\documentclass[10pt,tightenlines,twocolumn,prd,aps,floats,letterpaper,floatfix,nofootinbib,superscriptaddress,preprintnumbers]{revtex4}
\usepackage{amsmath,amssymb}
\usepackage{color}
\usepackage[dvips]{graphicx}
\definecolor{darkblue}{rgb}{0,0,0.5}
\usepackage[dvips, unicode, colorlinks, bookmarks=true, citecolor=darkblue, linkcolor=darkblue, urlcolor=darkblue]{hyperref}

\def\M{\mathcal{M}}

\begin{document}

\title{Optimizing the regimes of Advanced LIGO gravitational wave detector for multiple source types}

\author{I.S.~Kondrashov} 
\author{D.A.~Simakov}
\author{F.Ya.~Khalili}\email{farid@hbar.phys.msu.ru}
\affiliation{Physics Faculty, Moscow State University, Moscow 119992, Russia}
\author{S.L.~Danilishin}
\email{Stefan.Danilishin@aei.mpg.de}
\affiliation{Max-Planck-Institut f\"ur Gravitationsphysik (Albert-Einstein-Institut), Golm D-14482, Germany}
\affiliation{Physics Faculty, Moscow State University, Moscow 119992, Russia}

\date{\today}

\begin{abstract}
We develop here algorithms which allow to find regimes of signal-recycled Fabry-Perot--Michelson interferometer (for example, Advanced LIGO), optimized concurrently for two (binary inspirals + bursts) and three (binary inspirals + bursts + millisecond pulsars) types of gravitational waves sources. We show that there exists a relatevely large area in the interferometer parameters space where the detector sensitivity to the first two kinds of sources differs only by a few percent from the maximal ones for each kind of source. In particular, there exists a specific regime where this difference is $\approx0.5\%$ for both of them. Furthermore we show that even more multipurpose regimes are also possible, that provide significant sensitivity gain for millisecond pulsars with only minor sensitivity degradation for binary inspirals and bursts.
\end{abstract}

\maketitle

\section{Introduction}

Within the last decade we have witnessed a very significant progress in experimental gravitational wave (GW) astronomy. All the ground-based interferometric GW antennae such as LIGO \cite{06a1LIGO} in the USA, VIRGO \cite{06a1VIRGO} in Italy, GEO600 \cite{06a1GEO} in Germany and TAMA300 \cite{05a1TAMA} in Japan have been commissioned to operation and started to record scientific data. Nevertheless, no signs of gravitational waves were found thus far in this data which is, as we understand now, quite reasonable as it agrees with moderately optimistic predictions of the astrophysicists on the rate of measurable events within the limits of antennae detection range. 
This possibility was realized by GW community, and work on design of the next, second generation of GW antennae  went on in parallel with efforts in enhancement of the first generation ones. A pioneer amongst the second generation GW detectors will become an American Advanced LIGO project whose construction should start in 2010 \cite{Lazzarini2007}. It is planned to have sensitivity more than an order of magnitude higher than its predecessor. Such a dramatic increase will be provided by significantly lower seismic noise level due to new active antiseismic isolation, use of higher quality optics and lower level of quantum noise.

The main difference between the Initial LIGO and Advanced LIGO designs that is crucial for lowering this noise is the use of \textit{signal recycling} (SR) technique. Its implementation in contemporary detector setup is relatively easy as it requires to install only one additional mirror in the interferometer output port. This mirror reflects sideband signal field coming out of the interferometer back to the arm cavities or "recycles"  it. However, the dynamics and quantum noise properties of the interferometer become much richer and thus provide more freedom in adjustment its sensitivity curve to fit the current research goals. In particular, it was stressed by  A.~Buonnano and Y.~Chen \cite{02a1BuCh} that  the optical system composed of the SR cavity and the arm cavities forms a composite resonant cavity whose eigenfrequencies and quality factors can be controlled by the position and reflectivity of the SR mirror, thus increasing or decreasing the storage time of the signal inside the cavity. Moreover, as in Advanced LIGO it is planned to increase the optical power, circulating in the arm cavities approximately 80 times with respect to initial LIGO, an optomechanical interaction between laser field and mirrors will significantly influence the dynamics of the test masses turning them from free bodies (within the detection frequency band: $\sim 10 \div 10^4$ Hz) into oscillators with eigenfrequency falling into the detection band. This effect, known as ponderomotive rigidity \cite{70a1BrMaTi,02a1BuCh,99a1BrKh,01a2Kh,03a1eBiSa} arises when off-resonant optical field creates an effective restoring force originating from radiation pressure, which occurs to be a function of mirror displacement. In this situation one can say that optical field creates a frequency dependent mechanical rigidity \cite{01a1BuCh,01a2Kh,02a1BuCh,03a1BuCh,05a1LaVy,06a1KhLaVy}. 

As a result, the quantum noise spectral density of signal recycled interferometers (SRI) can be tuned to provide the best sensitivity for different gravitational wave sources. So far, it was supposed that in order to reach good sensitivity for each of the source types, totally different strategies should be used, which correspond to different optical parameters sets. Detection of gravitational waves from inspiraling neutron star binaries (NSNS) requires, for example, that the noise spectral density have to be as small as possible at low and medium frequencies, $f\lesssim 100\,{\rm Hz}$. If one is interested in narrow-band detection of GWs from the source with well defined center frequency $f_{\rm puls}$, such as high frequency pulsars, the optimal regime of the interferometer will be absolutely different: evidently, one should choose such set of optical parameters that provides minimum to noise at $f_{\rm puls}$. Searches of GWs from supernovae bursts or stochastic relic gravitational radiation which analytical waveforms are not known require flat broadband noise curves  \cite{04a1LSC-Bursts,05a1LSC-Bursts,07a1LSC-Bursts}.

\begin{figure}
  \begin{center}
 \includegraphics[width=.5\textwidth]{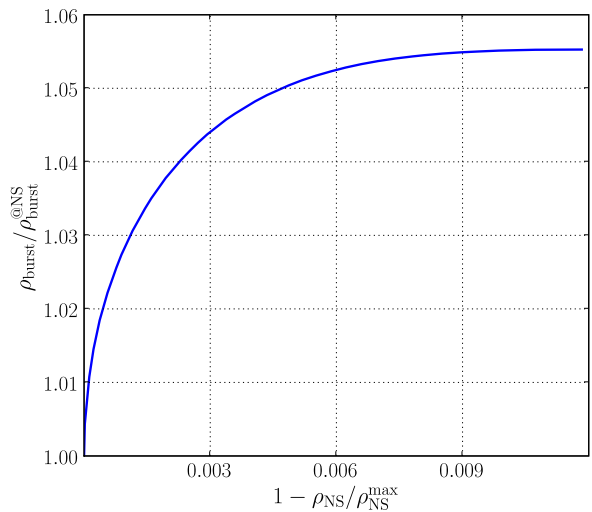}
\end{center}
  \caption{Relative improvement in Signal-to-Noise-Ratio $\rho_{\rm burst}/\rho_{\rm burst}^{\rm @NS}$ for bursts of GW radiation as a function of relative deterioration in SNR for neutron stars (NS) binaries $1-\rho_{\rm NS}/\rho_{\rm NS}^{\rm max}$. Here $\rho_{\rm burst}^{\rm @NS}$ is the value of SNR for GW burst sources when interferometer is tuned for reaching maximum SNR for NS binaries.}\label{fig0a}
 \end{figure} 
\begin{figure}
  \begin{center}
 \includegraphics[width=.5\textwidth]{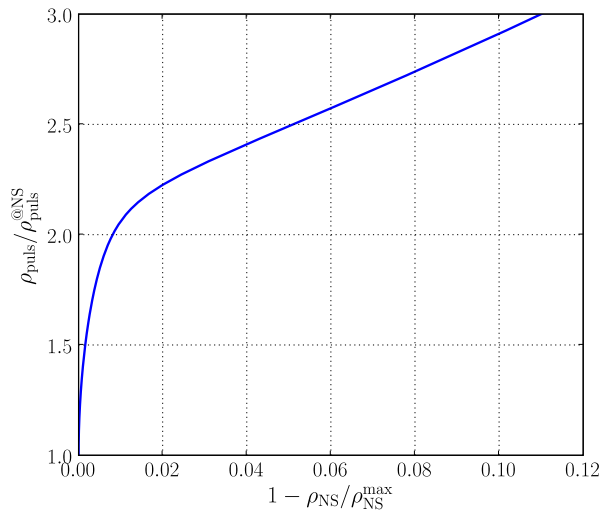}
\end{center}
\caption{Relative improvement in Signal-to-Noise-Ratio $\rho_{\rm puls}/\rho_{\rm puls}^{\rm @NS}$ for one of the high frequency pulsars (J0034-0534 with rotational frequency $f_0 \simeq 532,7$~Hz) as a function of relative deterioration in SNR for neutron stars (NS) binaries $1-\rho_{\rm NS}/\rho_{\rm NS}^{\rm max}$. Here $\rho_{\rm puls}^{\rm @NS}$ is the value of SNR for GWs from specified pulsar when interferometer is tuned for reaching maximum SNR for NS binaries.}\label{fig0b}
 \end{figure} 

 Extensive studies of optimal modes of operation of Advanced LIGO interferometer most suitable for different specific kinds of sources of GWs were carried out in AdvLIGO Lab in extensive details \cite{08tnLIGOlab}. However, in this paper we focus on finding multipurpose regimes of signal recycled interferometers which might provide good (sub-optimal) sensitivity simultaneously for different gravitational wave source types. The possibility to do this originates from the fact that classical noise budget for currently operating and future GW detectors masks low and medium frequency features of the quantum noise (see Fig. \ref{fig3}) that are mostly susceptible to variation of optical parameters of the interferometer. This fact, although being quite unpleasant for tunability of GW detector for specific sources is surprisingly advantageous for tuning the antenna to have high enough sensitivity to GWs from various types of astrophysical sources \textit{simultaneously}. As will be shown in subsequent sections the sensitivity changes relatively slowly within quite a wide range of main SRI optical  parameters for different types of signals (GWs from inspiraling compact binaries, GW bursts, high-frequency pulsars etc.).
 And these areas for different sources significantly overlap that allows to find quasi-optimal regime for two or even three different GW sources simultaneously. It is shown by the example of Advanced LIGO SRI that rather significant improvement in sensitivity to GW bursts and GWs from high frequency sources is possible at the cost of quite moderate deterioration of signal strength for compact binary systems. This is illustrated in Figs. \ref{fig0a}~and~\ref{fig0b} where the relative improvement in Signal-to-Noise-Ratio (SNR) for GW bursts and high frequency pulsars, correspondingly, are plotted with respect to relative deterioration of SNR for NSNS, provided that one diverts optical parameters of SRI from the optimal ones for NSNS. It should be also emphasized here that in spite Advanced LIGO is used as an example for which we perform calculations, the results we obtain are general and applicable to all SRI limited by classical noise at low and medium frequencies. It is also instructive to mention that our optimization includes only the most basic parameters and the results can be considered only as some preliminary guidelines for designing future generation of GW interferometers, while for more specific optimization of real device much larger parameter space should be considered and optimization over parameters that should be set before the device is built and the ones, that can be easily tuned in already operating detector should be performed separately. The above issues will be considered in future works.

The paper is organized as follows. In Section \ref{Sec1} the brief consideration of quantum noise of signal-recycled Fabry-P\'erot--Michelson interferometer is performed. In Section \ref{Sec2} expressions for Signal-to-Noise Ratio  and detection range for the gravitational-wave radiation from the inspiraling binary system are given and the numerical optimization procedure with respect to interferometer optical parameters is described. The quantitative and qualitative analysis of the obtained results of optimization against GWs from neutron star binaries is performed. 
In Section \ref{Sec3} the sensitivity of interferometer to GW bursts is analyzed and regimes for simultaneous detection of these two types of sources are investigated. In section \ref{Sec4} this analysis extended also to high-frequency quasi-periodic sources (pulsars). In Section \ref{Conc} the obtained results are discussed and some concluding remarks are given. Some notations and values of parameters used through this paper are listed in Table\,\ref{tab1}.

\begin{table}[]
 \begin{tabular}{|c|l|}
 \hline
Value & \multicolumn{1}{|c|}{Description}\\
\hline
$M$ & Test bodies reduced mass\\
$c$ & Speed of light\\
$L$ & SRI arms length\\
$\omega_p$ & Laser frequency\\
$\Omega$ & Mechanical frequency\\
$W$ & Circulating optical power\\
$\gamma$ & Effective SRI half-bandwidth\\
$\delta$ & Effective SRI detuning\\
$J=\dfrac{4\omega_pW}{McL}$ & Renormalized optical power\\
$\phi$ & Homodyne angle \\
$\eta$ & Total readout quantum efficiency (incl. losses)\\
\hline
\end{tabular} 
\caption{Notations used for characterizing quantum noise of SRI}\label{tab1}
\end{table}

\section{Quantum noise of signal recycled interferometers}\label{Sec1}

\begin{figure}
  \begin{center}
 \includegraphics[width=.5\textwidth]{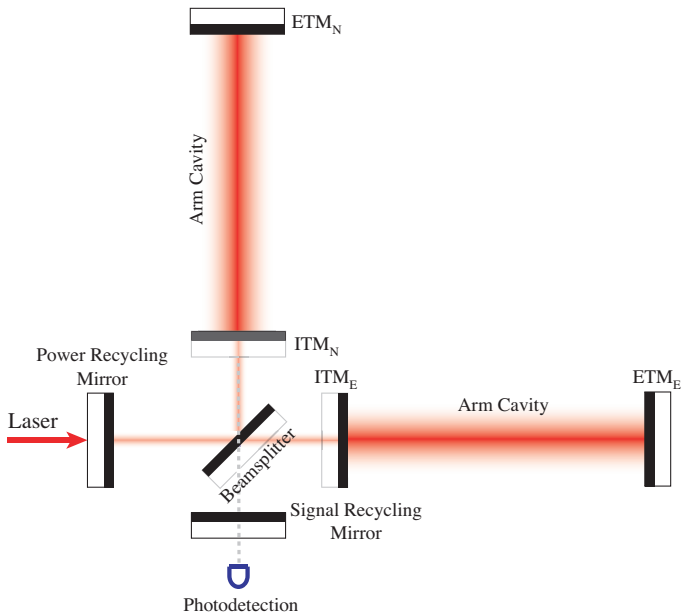}
\end{center}
  \caption{Principle optical scheme of signal-recycled interferometer of Advanced LIGO GW detector}\label{fig1}
 \end{figure} 
 
 In Fig.\,\ref{fig1} the schematic drawing of a signal recycled interferometer is presented. Here the additional signal recycling mirror (SRM) forms, together with the input test masses (ITMs) of arm cavities, an additional SR cavity which properties are defined by two parameters of SRM, namely its amplitude reflectivity $\rho$ and detuning phase $\phi_{\rm SRC} = [\omega_pl/c]_{\!\!\!\mod{2\pi}}$ 
gained by carrier light travelling one-way in the SR cavity ($l$ is the length of SR cavity). 

As demonstrated by A.~Buonnano and Y.~Chen
\cite{03a1BuCh}, there exists one-to-one transformation (``scaling law'') between the parameters of a single detuned Fabry-Perot (FP) cavity with one movable mirror 
and ones of SRI that allows to describe the optical behavior of it in terms of much more simple equivalent system such as FP cavity. According to ``scaling law'' for any SRI there exists a unique FP cavity with bandwidth $\gamma$ and detuning $\delta$ defined by formulae:
\begin{subequations}\label{ScLaw}
  \begin{gather}
    \gamma = \dfrac{(1-\rho^2)\gamma_{\rm ARM}}{1+2\rho\cos2\phi_{\rm SRC}+\rho^2}\,, \\
    \delta 
      = \dfrac{2\rho\gamma_{\rm ARM}\sin{2\phi_{\rm SRC}}}{1+2\rho\cos2\phi_{\rm SRC}+\rho^2}\,,
  \end{gather}
\end{subequations}
where $\gamma_{\rm ARM}=cT/4L$ is the half-bandwidth of arm FP cavities, that has the same optomechanical features and therefore the same sensitivity as the initial SRI. The effective optical power circulating in the equivalent FP cavity should be twice as large as real optical power $W$ circulating in a single arm cavity. The same is referred to masses of input (ITM) and end (ETM) test masses of effective cavity: $M_{\rm eff}=2M$.

Below we will use also extensively the following convenient parameters: generalized bandwidth
\begin{subequations}
  \begin{gather}
    \varGamma = \sqrt{\gamma^2+\delta^2} = \gamma_{\rm ARM}
      \sqrt{\frac{1-2\rho\cos2\phi_{\rm SRC}+\rho^2}{1+2\rho\cos2\phi_{\rm SRC}+\rho^2}} 
    \intertext{and detuning phase}
    \beta = \arctan\frac{\delta}{\gamma} 
      = \arctan\left(\frac{2\rho}{1-\rho^2}\,\sin{2\phi_{\rm SRC}}\right) \,.
  \end{gather}
\end{subequations}
The above expressions can be also easily reverted to obtain the SR cavity parameters:
\begin{subequations}
  \begin{gather}
    \rho = \sqrt{
      \frac{\gamma_{\rm ARM}^2 - 2\gamma_{\rm ARM}\varGamma\cos\beta + \varGamma^2}
        {\gamma_{\rm ARM}^2 + 2\gamma_{\rm ARM}\varGamma\cos\beta + \varGamma^2}
      } \,, \\
    \phi_{\rm SRC} = \frac{1}{2}\begin{cases}
      \arcsin\left(\dfrac{1-\rho^2}{2\rho}\,\tan\beta\right)\,, & \varGamma<\gamma_{\rm ARM} \,, \\
      \pi - \arcsin\left(\dfrac{1-\rho^2}{2\rho}\,\tan\beta\right)\,,& \varGamma>\gamma_{\rm ARM}\,.
    \end{cases}
  \end{gather}
\end{subequations}

Using the ``scaling law'' approach, consider FP cavity with movable mirrors pumped by laser light with frequency $\omega_0$. Action of gravitational wave on such a system can be effectively described by means of effective forces acting on the mirrors and therefore changing dynamically the phase shift of outgoing light with respect to ingoing one. 

There exist two kinds of quantum fluctuations limiting the sensitivity of detector. They are so called laser shot noise (SN) and radiation pressure noise (RPN). The first one originates from quantum fluctuations of electromagnetic wave phase which prevents from exact phase shift measurement and is, in essence, measurement accuracy. The second one, being a consequence of fluctuations of light amplitude, causes random radiation pressure force to move the mirrors and masks the measured signal force. As far as this noise is the direct consequence of measurement, it is also known as back action noise because of the back action of the measurement device (laser light) on measured quantity (phase shift). 

\begin{figure}
 \centering
 \includegraphics[width=.5\textwidth]{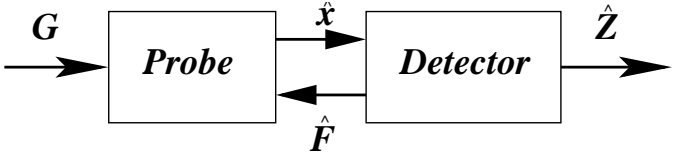}
 \caption{Schematic diagram of the quantum measurement device. $G$ is the classical observable (\textit{e.g.} force) acting on the probe that is measured. $\hat Z$ is the output signal of the detector device, $\hat x$ is the linear observable (\textit{e.g.} displacement) of the probe, and $\hat F$ is the linear observable of the detector which describes the back-action force on the probe.}
 \label{fig2}
\end{figure}

It is convenient to describe this system in terms of linear quantum measurement theory formalism developed in \cite{92BookBrKh} and thoroughly elaborated for use in gravitational-wave interferometers in \cite{02a1BuCh}. Following this formalism we can represent our meter as two linearly coupled systems, probe (test masses) and detector (laser light and photodetectors).  The schematic drawing of this equivalent linear system is presented in Fig. \ref{fig2}. Here $\hat x$ stands for some measured observable of the probe (mirrors relative displacement in our case), $\hat F$ is some observable of the detector through which it is coupled to the probe (radiation pressure force in our case), $G$ is classical signal force being detected (GW action on the detector) and $\hat Z$ is the measured observable of the detector (output light quadrature in our case). Following \cite{03a1BuCh} we write down the Hamiltonian of our system as:
\begin{equation}
 \hat H = \bigl[(\hat H_{\cal P}-\hat xG)+\hat H_{\cal D}\bigr]-\hat x\hat F \equiv \hat H^{(0)}+\hat V\,,
\end{equation}
where $\hat H^{(0)}=(\hat H_{\cal P}-\hat xG)+\hat H_{\cal D}$ is considered as zeroth order Hamiltonian for both detector (marked by ${\cal D}$) and probe  (marked by ${\cal P}$) subsystems, and linear coupling between them $\hat V = -\hat x\hat F$ considered as a perturbative Hamiltonian. 
Using this approach, one can write down all of the observables of the system as a sum of unperturbed zeroth order terms (marked by subscript $^{(0)}$) and perturbations (marked by subscript $^{(1)}$) (see Eqs. (2.12-2.14) of \cite{03a1BuCh}). In frequency domain these observables are read as:
\begin{subequations}
  \begin{gather}
    \hat Z^{(1)}(\Omega) = \hat Z^{(0)}(\Omega) + R_{ZF}(\Omega)\hat x^{(1)}(\Omega)\,,\\
    \hat F^{(1)}(\Omega) = \hat F^{(0)}(\Omega) + R_{FF}(\Omega)\hat x^{(1)}(\Omega)\,,\\
    \hat x^{(1)}(\Omega) = \hat x^{(0)}(\Omega) + Lh(\Omega)/2 
      + R_{xx}(\Omega)\hat F^{(1)}(\Omega)\,.
  \end{gather}
\end{subequations}
Here $Lh(\Omega) \equiv R_{xx}G(\Omega)$ is the GW signal proportional to metrics variation spectrum $h(\Omega)$. Quantities $R_{AB}(\Omega)$ are frequency-dependent susceptibilities, in particular, $R_{ZF}(\Omega)$ is the optomechanical coupling factor,
\begin{equation} R_{xx}(\Omega) = -\dfrac{1}{M\Omega^2} \end{equation}
is mechanical susceptibility of the SRI and
\begin{equation}
 R_{FF}(\Omega) = \dfrac{MJ\delta}{\varGamma^2 - \Omega^2 - 2i\gamma\Omega}\,,
\end{equation}
is the optical rigidity. $\hat Z^{(0)}$ corresponds to shot noise of the laser light, $x^{(0)}\equiv x_{\rm tech}$ stands for any displacement noise sources associated with the test mass reflecting surface with respect to its center of mass position, namely
thermoelastic and Brownian noise of the mirror coatings and substrate, and
\begin{equation}
  \hat F^{(0)}(\Omega) = \hat F_{\rm RPN}(\Omega) + F_{\rm tech}(\Omega) \,,
\end{equation}
where $\hat F_{\rm RPN}(\Omega)$ is the radiation pressure noise and $F_{\rm tech}(\Omega)$ is describing all the classical force noises, most notably suspension thermal, gravity gradient and seismic noises.

Using these definitions one can now write down the output of the SRI reduced to metrics variation $h$ units as: 
\begin{multline}
  \hat h_{\rm out}(\Omega) = h(\Omega) + \frac{2}{L}\Bigl\{
    R_{xx}(\Omega)[\hat F_{\rm RPN}(\Omega)  + F_{\rm tech}(\Omega)] \\
    + [R_{xx}(\Omega)R_{FF} + 1][\hat x_{\rm SN}(\Omega) + x_{\rm tech}(\Omega)] 
  \Bigr\} \,,
\end{multline}
where
\begin{equation}\label{x_fluct}
  \hat x_{\rm SN}(\Omega) = \dfrac{\hat Z^{(0)}}{R_{ZF}(\Omega)} 
\end{equation}
is the normalized shot noise.

Accounting for these definitions, one can write down now spectral density of the interferometer output noise as:
\begin{equation}\label{SRIoutSD}
  S^h(\Omega) = S^h_{\rm quant}(\Omega) + S^h_{\rm tech}(\Omega) \,.
\end{equation}
Here
\begin{multline}
  S^h_{\rm quant}(\Omega) = \frac{4}{L^2}\Bigl(
    |R_{xx}(\Omega)R_{FF}(\Omega)+1|^2S_x^{\rm SN}(\Omega) \\
    + 2\Re\{[R_{xx}(\Omega)R_{FF}(\Omega)+1]^*S_{xF}(\Omega)\}
    + |R_{xx}(\Omega)|^2S_F^{\rm RPN}(\Omega)
  \Bigr)
\end{multline}
is the sum quantum noise spectral density,
\begin{equation}\label{S_x}
  S_x^{\rm SN}(\Omega) = \dfrac{\hbar}{2MJ\gamma\eta}
    \dfrac{\Omega^4 + 2\varGamma^2\Omega^2\cos2\beta + \varGamma^4}
      {\varGamma^2\cos^2(\beta+\phi) + \Omega^2\cos^2\phi}
\end{equation}
is the shot noise (\ref{x_fluct}) spectral density, 
\begin{equation}
 S_F^{\rm RPN}(\Omega) = \dfrac{2\hbar MJ\gamma(\varGamma^2 + \Omega^2)}
  {\Omega^4 + 2\varGamma^2\Omega^2\cos2\beta + \varGamma^4} \,,
\end{equation}	
is radiation pressure noise $\hat F_{\rm RPN}(\Omega)$ spectral density, 
\begin{equation}  
  S_{xF}(\Omega) = \hbar\,\dfrac{\varGamma\sin(\beta+\phi) + i\Omega\sin\phi}
    {\varGamma\cos(\beta+\phi) + i\Omega\cos\phi}
\end{equation}
is their spectral cross-correlation function,
\begin{multline}
  S^h_{\rm tech}(\Omega) = \frac{4}{L^2}\Bigl(
    |R_{xx}(\Omega)R_{FF}(\Omega)+1|^2S_x^{\rm tech}(\Omega) \\
    + |R_{xx}(\Omega)|^2S_F^{\rm tech}(\Omega) 
  \Bigr)
\end{multline}
is the sum technical noise spectral density, and $S_F^{\rm tech}(\Omega)$ and $S_x^{\rm tech}(\Omega)$ are spectral densities of non-quantum noise sources  $F_{\rm tech}(\Omega)$ and $x_{\rm tech}(\Omega)$. Optical losses influence, as shown in \cite{06a1KhLaVy}, can be accounted for by introducing effective quantum efficiency  $\eta$ of the readout photodetector, which appears in Eq.\,(\ref{S_x}).

\section{Binary sources}\label{Sec2}

The most popular and easy to implement criteria used to determine the optimal regime of GW detectors relates to evaluation of detection range for inspiraling binary systems of compact objects such as neutron stars and/or black holes. This method is based on estimation of Signal-to-Noise-Ratio (SNR) using well known analytical expression for spectral density of GWs emitted by system of two gravitationally bounded inspiraling astrophysical objects (See Sec.~3.1.3 of \cite{06a1PoYu}):
\begin{equation}\label{GWNS}
  |h(f)|^2 = \frac{G^{5/3}}{c^3} \frac{\pi}{12} \frac{\M^{5/3}}{r^2}
  \frac{\Theta(f_{\rm max}-f)}{(\pi f)^{7/3}} \,,
\end{equation}
where  $\M\equiv\mu^{3/5}M^{2/5}$ is the so called "chirp mass" of the binary system constructed from reduced mass $\mu=M_1M_2/M$ and total mass $M=M_1+M_2$ of the binary system with components masses $M_1$ and $M_2$ correspondingly. One can readily see the indicative frequency dependence $|h(f)|^2\propto f^{-7/3}$ and inverse dependence on distance to the system squared $r^2$. The upper cutoff frequency $f_{\rm max}$ corresponds to the period of binary system rotation on innermost stable circular orbit (ISCO) when the system goes from quasistationary rotation phase to non-stationary merger phase. This frequency can be estimated as:
 \begin{equation}\label{f_max}
  f_{\rm max}\simeq 4400\times(M_\odot/M)\,\mbox{Hz}\,.
 \end{equation}

Given GW signal shape (\ref{GWNS}) and the noise spectral density (\ref{SRIoutSD}), it is possible to write down the optimal SNR $\rho$ which can be obtained on a given detector. As demonstrated by E.~Flanagan and S.~Hughes \cite{98a1FlHu}, SNR averaged over all mutual orientations between the detector and source and over both polarizations of GWs is equal to:
\begin{multline}
  \rho_{\rm NS}^2
  = \dfrac{4}{5}\int_{f_{\rm min}}^{f_{\rm max}}\dfrac{|h(f)|^2}{S^h(2\pi f)}\,df \\
  = \dfrac{2}{15}\dfrac{G^{5/3}}{\pi^{4/3}c^3}\dfrac{\M^{5/3}}{r^2}
      \int_{f_{\rm min}}^{f_{\rm max}}\dfrac{df}{f^{7/3}S^h(2\pi f)}\,,
\label{SNR_inspiral}
\end{multline}
where $f_{\rm min}$ is the lower cutoff frequency at which binary system motion cannot be considered as stationary. In our calculations we will take $f_{\rm min}\simeq 10$ Hz.
 
In order to estimate detection range $r$ one should set a threshold SNR $\rho_0$ which defines the level of confidence in detection of GWs from binary system. Then detection range can be written as
\begin{equation}\label{range}
  r = \left(\dfrac{2}{15}\dfrac{G^{5/3}}{\pi^{4/3}c^3}\dfrac{\M^{5/3}}{\rho_0^2}
    \int_{f_{\rm min}}^{f_{\rm max}}\dfrac{df}{f^{7/3}S^h(f)}\right)^{1/2}\,.
\end{equation}
 
Of course, all formulae we use here are obtained in the lowest Post-Newtonian order of general relativity \cite{00a1DaIySa} that definitely limits their application area to stellar masses systems and asymptotically flat space-time background. However, for our purposes it is enough and the most significant for us feature of the above expressions is their relative simplicity. 
 
\begin{figure}[t]
  \includegraphics[width=.49\textwidth]{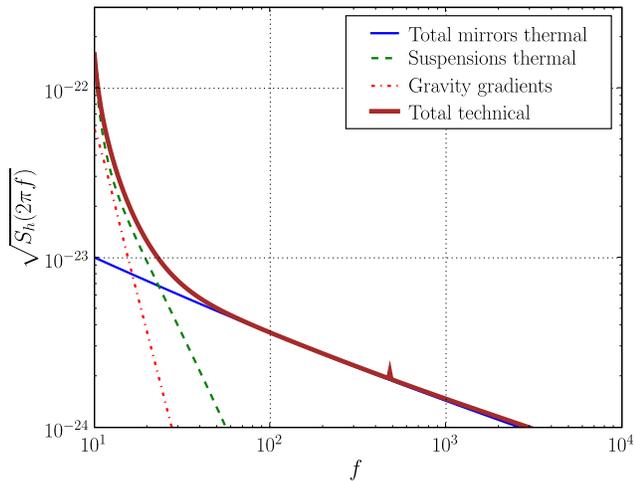}
  \caption{Main classical noises planned for Advanced LIGO interferometer.}\label{fig3}
\end{figure}

We calculated the detection range for standard ($M = 2.8M_\odot$) neutron stars binary system numerically considering it as a function of 3 parameters: $\varGamma\in[500,\,12500]\,{\rm s}^{-1}$, $\beta\in[-\pi/2,\,\pi/2]$, and $\phi\in[-\pi/2,\,\pi/2]$. The number of values for each of the parameters was 193, giving in total $193^3\approx7.2\cdot10^6$ optical configurations. Distribution of points in parameter space was taken uniform over angle variables $\beta$  and $\phi$, and logarithmic over variable $\varGamma$. For other parameters, we used the values planned for Advanced LIGO, see Table\,\ref{tab1}. To account for technical noises, we used the noise budget also planned for Advanced LIGO and generated by {\sc bench} software \cite{bench}, see in Fig.\,\ref{fig3}. 

\begin{figure}[t]
  \includegraphics[width=.49\textwidth]{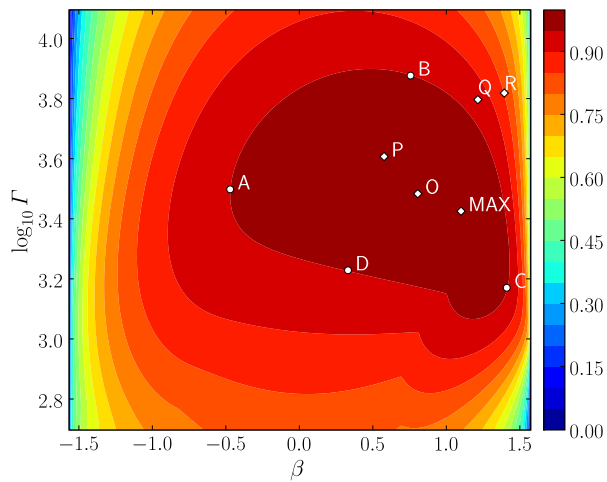}
  \caption{
Contour plot of sensitivity for standard NS-NS binaries $\rho_{\rm NSNS}/\rho_{\rm NSNS}^{\rm max}$  as a function of $\Gamma,\,\beta$. Point ``MAX'' corresponds to the sensitivity maximum, points ``A''---`D'' --- to typical suboptimal tunings shown in Fig.\,\ref{fig4.5}, point ``O'' --- to double (NSNS+bursts) optimal regime, and points ``P'',``Q'',``R'' --- to triple (NSNS+bursts+pulsar) suboptimal regimes. 
}\label{fig4}
 \end{figure}

For each pair of $\varGamma,\,\beta$ we have maximized SNR with respect to $\phi$, thus obtaining a function of only two parameters $\varGamma,\,\beta$. 
The result of this calculation is presented in Fig.~\ref{fig4} as 
contour plots of normalized SNR $\rho_{\rm NSNS}/\rho_{\rm NSNS}^{\rm max}$. Contours act in this plot as margins for regions in parameter space where SNR is higher than the certain percentage of maximal SNR $\rho_{\rm NSNS}^{\rm max}$, being indicated in plot by point marked as ``MAX''. The parameters values for this maximal sensitivity point are listed in Table\,\ref{tab4.5}.

It is easy to note flat behaviour of SNR within a spacious range of parameters $\varGamma$, $\beta$. It arises due to two reasons. The first one is technical noises. It can be shown that in the absence of them, it is possible, in principle, to obtain arbitrary high values of SNR using deep and narrow well in quantum noise spectral density created by means of 
the second-order-pole regime of the optical rigidity \cite{01a2Kh, 06a1KhLaVy}, which corresponds to $\delta=(J/4)^{1/3}$ and $\gamma\to0$. Technical noise which has flat and smooth spectral dependence makes such excesses in the quantum noise useless. Moreover, they increase the quantum noise at other frequencies and thus decrease the sensitivity. 

\begin{figure}[t]
  \begin{center} 
    \includegraphics[width=.49\textwidth]{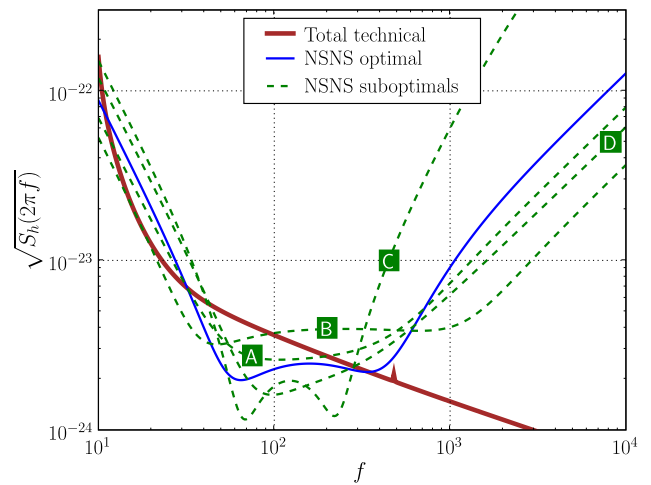}
  \end{center}
  \caption{Quantum noise spectral densities optimized for standard NSNS binary sources (point ``MAX'' in Fig.\,\ref{fig4}) and four typical sub-optimal spectral densities (points ``A''---``D'' (see parameters in Table.~\ref{tab5.5}) in Fig.\,\ref{fig4}).}\label{fig4.5}
\end{figure}

On the other hand, the integral character of criterion (\ref{range}) allows significant variations of values of $\varGamma$ and $\beta$. Increase of $\varGamma$ decreases quantum noise level at low frequencies ($f\lesssim100\,{\rm Hz}$), but increases it at medium frequencies ($100\,{\rm Hz}\lesssim f\lesssim1000\,{\rm Hz}$), and {\it vice versa}. Moderate positive values of detuning $\delta$ (and hence positive values of $\beta$) create resonance ``wells'' in the noise spectral density, but increase the low-frequency noise. As a result, the integral (\ref{range}) does not change significantly.

This consideration can be illustrated by Fig.\,\ref{fig4.5}, where quantum noise spectral densities  for the optimal parameters set (point ``MAX'' in Fig.\,\ref{fig4}) as well as for 4 typical sub-optimal ones (points ``A''---``D'') are plotted.

\begin{table}
  \begin{tabular}{|c|c|c|cccc|}
    \hline                      & NSNS & Bursts & \multicolumn{4}{|c|}{NSNS+Bursts+Pulsar} \\
    \hline                      & MAX  & MAX    & ``O'' & ``P'' & ``Q'' & ``R''\\
    \hline
     $\varGamma$                & 2700 & 2900   & 3100 & 4600 & 6400 & 6600 \\
     $\beta$                    & 1.10 & 0.57   & 0.80 & 0.52 & 1.19 & 1.39 \\
     $\phi$                     &-1.00 &-0.23   &-0.44 &-0.23 &-0.85 &-1.10 \\
     $\rho^2$                   & 0.84 & 0.74   & 0.79 & 0.80 & 0.94 & 0.97 \\
     $\phi_{\rm SRC}$           & 1.48 & 1.52   & 1.51 & 1.54 & 1.53 & 1.53 \\
     $\gamma=\varGamma\cos\beta$& 1200 & 2400   & 2200 & 4000 & 2400 & 1200 \\
     $\delta=\varGamma\sin\beta$& 2400 & 1600   & 2200 & 2300 & 5900 & 6500 \\
     $\rho_{\rm NS}/\rho_{\rm NS}^{\rm max}$       & 1.0    & 0.988 & 0.995 & 0.989 & 0.91 & 0.84 \\
     $\rho_{\rm burst}/\rho_{\rm burst}^{\rm max}$ & 0.979  & 1.0   & 0.995 & 0.989 & 0.84 & 0.75 \\
     $\rho_{\rm puls}/\rho_{\rm puls}^{\rm max}$   & 0.21   & 0.26  & 0.25  & 0.30  & 0.40 & 0.50 \\
    \hline
  \end{tabular} 
  \caption{Optimal (``MAX''), double-optimal (``O''), and triple-suboptimal (``P'',``Q'',``R'') parameters values for standard NSNS binary, burst and typical periodic (pulsar J0034-0534) sources. }\label{tab4.5}
\end{table}

\begin{table}
 \begin{tabular}{|c|cccc|cccc|}
 \hline                      & \multicolumn{4}{|c|}{NSNS} & \multicolumn{4}{|c|}{Bursts}\\
    \hline                        & ``A'' & ``B'' & ``C'' & ``D'' & ``A'' & ``B'' & ``C'' & ``D''\\
    \hline
$\varGamma$             & 3161  & 7559 & 1487 & 1700 & 3057 & 6182 & 2010 & 1438 \\
$\beta$                    &-0.47 & 0.75&  1.41 & 0.33 & -0.70 & 0.49 & 1.24 & 0.20\\
$\phi$			&0.97 &-0.49 & 1.49 & 0.18 & 0.47 & -0.29 & -0.57 & 0.03\\
$\rho^2$			&0.85 & 0.97 & 0.73 & 0.75 & 0.84 & 0.95 & 0.77 & 0.73\\
$\phi_{\rm SRC}$	&1.55 & 1.54 & 0.09 & 0.06 & 1.55 & 1.54 & 0.03 & 0.10\\
\hline
 \end{tabular}
\caption{Optical parameters for the SRI having quantum noise plotted in Fig.~\ref{fig4} and Fig.~\ref{fig5.5}.}\label{tab5.5}
\end{table}

\section{Gravitational-waves bursts}\label{Sec3}

The next type of possible GW sources are supernovae explosions and stellar cores collapses, compact binary systems mergers \cite{06a1BaCeChKoMe} and other sources with not well modeled properties which are usually called GW bursts \cite{04a1LSC-Bursts,05a1LSC-Bursts,07a1LSC-Bursts}. For these sources, we use the simple model of logarithmic-flat signal spectrum over the range of frequencies from $f_l$ to $f_h$. By logarithmic-flat spectrum we mean that spectrum of GW signal $h(f)$ is proportional to $f^{-1/2}$, that corresponds to constant numerator in expression for SNR if integration is performed with respect to $\log f$:
\begin{equation*}
  \rho_{\rm burst}^2 \propto \int_{f_l}^{f_h}\frac{|h(f)|^2df}{S^h(2\pi f)}
  \propto \int_{\log f_l}^{\log f_h}\frac{d\log f}{S^h(2\pi f)} \,.
\end{equation*}
This way of defining SNR for burst events seems reasonable
to characterize astrophysical signals with unknown spectrum structure so that contributions from frequency (time) ranges of different order are equal (for example, contributions from $10\div100$~Hz and $100\div1000$~Hz should be the same).

\begin{figure}[t]
 \begin{center} 
 \includegraphics[width=.49\textwidth]{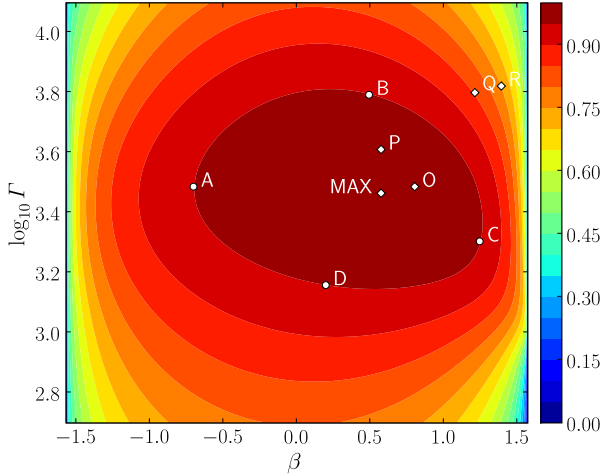}
\end{center}
\caption{
Contour plot of bursts sensitivity $\rho_{\rm burst}/\rho_{\rm burst}^{\rm max}$  as a function of $\Gamma,\,\beta$. Point ``MAX'' corresponds to the sensitivity maximum, points ``A''---``D'' (see parameters in Table.~\ref{tab5.5}) --- to typical suboptimal tunings shown in Fig.\,\ref{fig5.5}, point ``O'' --- to double (NSNS+bursts) optimal regime, and points ``P'',``Q'',``R'' --- to triple (NSNS+bursts+pulsar) suboptimal regimes.
}\label{fig5}
 \end{figure}

\begin{figure}[t]
  \begin{center} 
    \includegraphics[width=.49\textwidth]{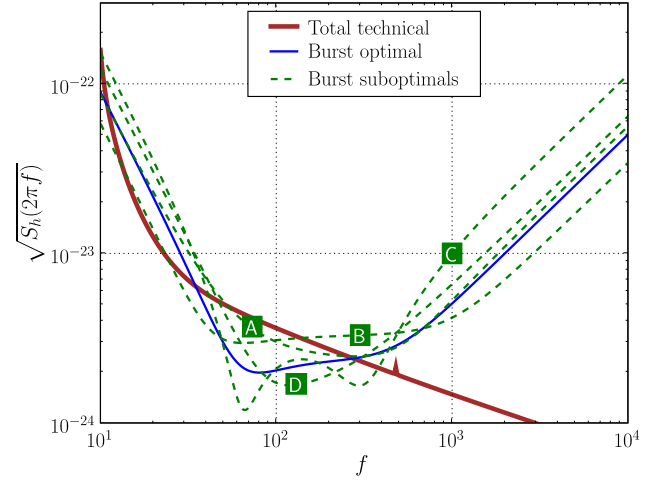}
  \end{center}
  \caption{Quantum noise spectral densities optimized for burst sources (point ``MAX'' in Fig.\,\ref{fig5}) and four typical sub-optimal spectral densities (points ``A''---``D'' in Fig.\,\ref{fig5} with parameters given in  Table.~\ref{tab5.5}).}\label{fig5.5}
\end{figure}

In Fig.\,\ref{fig5}, function $\rho_{\rm burst}(\varGamma,\beta)$, calculated using the same algorithm as in the previous (NSNS) case, and normalized by its maximal value $\rho_{\rm burst}^{\rm max}$, is shown. Similar to the previous case, in Fig.\,\ref{fig5.5} quantum noise spectral densities for the optimal parameters set (point ``MAX'' in Fig.\,\ref{fig5} and for 4 sub-optimal ones (points ``A''---``D'') are plotted, and parameters values for the point ``MAX'' are listed in Table\,\ref{tab4.5}.

It follows from Fig.\,\ref{fig5}, that function $\rho_{\rm burst}(\varGamma,\beta)$ also has flat behavior within almost the same range of $\varGamma$, $\beta$ as $\rho_{\rm NS}(\varGamma,\beta)$. The main difference between the NSNS and burst cases, invisible in Figs.\,\ref{fig4} and \ref{fig5}, stems from the existence of cut-off frequency (\ref{f_max}) and from more steep frequency dependence of the NSNS signal. As a result, the NSNS optimization procedure leads to smaller values of angle $\phi+\beta$, that reduces quantum noise at low frequencies, while the optimization with respect to GW bursts requires smaller values of $\phi$ that reduces quantum noise at high frequencies (compare Figs.\,\ref{fig4.5}, \ref{fig5.5} and the corresponding columns in Table \ref{fig4.5}).

However, this difference is quite subtle, and it is evident that regimes have to exist which provide good sensitivity for both these types of GW sources simultaneously. In order to find them, we calculate values of $\varGamma$, $\beta$, $\phi$, which maximize the combined normalized sensitivity 
\begin{equation}
  G_{\rm NS+burst}(\lambda)
  = \lambda\left(\frac{\rho_{\rm NS}}{\rho_{\rm NS}^{\rm max}}\right)^2
    + (1-\lambda)\left(\frac{\rho_{\rm burst}}{\rho_{\rm burst}^{\rm max}}\right)^2 ,
\end{equation}
where $\lambda$ is a Lagrange multiplier which varies in the range $[0,1]$. 

\begin{figure}[t]
  \begin{center} 
    \includegraphics[width=.49\textwidth]{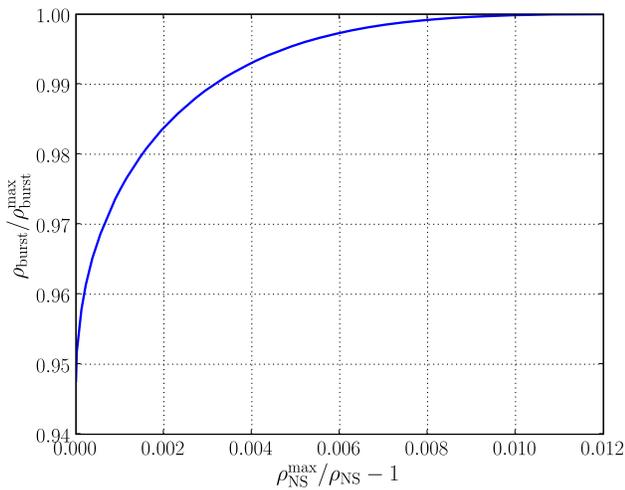}
  \end{center}
  \caption{Parametric plot of $\rho_{\rm burst}(\lambda)/\rho_{\rm burst}^{\rm max}$ against $\rho_{\rm NS}(\lambda)/\rho_{\rm NS}^{\rm max}$}\label{fig:nsns_burst}
\end{figure}

The result is shown in Fig.\,\ref{fig:nsns_burst}, where parametric plot of $\rho_{\rm burst}(\lambda)/\rho_{\rm burst}^{\rm max}$ against $\rho_{\rm NS}(\lambda)/\rho_{\rm NS}^{\rm max}$ is presented. The leftmost point on this plot corresponds to $\lambda=1$, and the rightmost one --- to $\lambda=0$. It follows from this calculation, that the tuning exists where values of $\rho_{\rm NS}$ and $\rho_{\rm burst}$ decrease both only by $\approx 0.5\%$ compared to their maximal values:
\begin{equation}
  \frac{\rho_{\rm NS}}{\rho_{\rm NS}^{\rm max}} \approx
    \frac{\rho_{\rm burst}}{\rho_{\rm burst}^{\rm max}} \approx 0.995 \,.
\end{equation}
The corresponding values of parameters $\varGamma$, $\beta$, and $\phi$ are listed in the column ``O'' of Table\,\ref{tab4.5}. This point on the plane \{$\varGamma$, $\beta$\}, marked as ``O'', is shown also in plots \ref{fig4}, \ref{fig5}.

\begin{figure}[t]
  \begin{center} 
    \includegraphics[width=.49\textwidth]{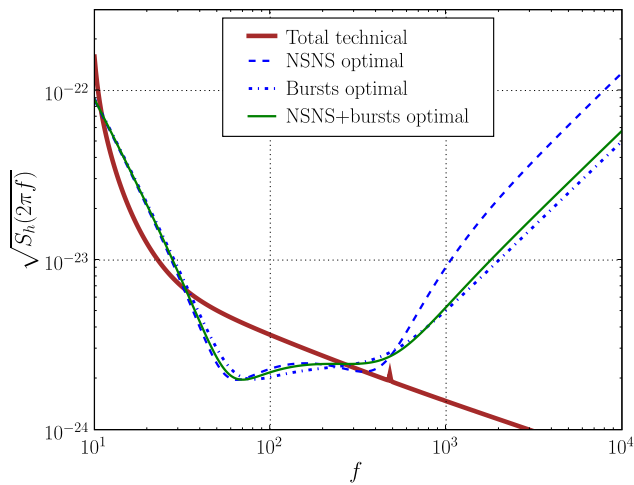}
  \end{center}
  \caption{Quantum noise spectral densities optimized for NSNS sources (point ``MAX'' in Fig.\,\ref{fig4}), bursts sources (point ``MAX'' in Fig.\,\ref{fig5}) and for both of them (point ``O'' in Figs.\,\ref{fig4},\ref{fig5}).}\label{fig:995}
\end{figure}

In Fig.\,\ref{fig:995} the plot of quantum noise spectral density for this regime is presented together with plots of previously calculated optimal spectral densities for NSNS and burst sources. 
Note that all three curves are virtually indistinguishable at low and medium frequencies ($f\lesssim600\,{\rm Hz}$), and the ones for bursts and combined NSNS+bursts are almost the same over all frequency ranges of interest.

\section{High-frequency periodical sources of GWs}\label{Sec4}

High frequency periodical sources of GWs, namely millisecond pulsars, can be treated as very narrow-band almost monochromatic sources with well defined central frequency $2f_{\rm puls}$ \cite{98a1JaKrSch}. For these sources, the detection range and the SNR are simply proportional to inverted square root of the noise spectral density at given frequency $2f_{\rm puls}$,
\begin{equation}
  r_{\rm puls} \propto \rho_{\rm puls} \propto \frac{1}{\sqrt{S^h(4\pi f_{\rm puls})}} \,.
\end{equation}
Direct optimization of quantum noise in this case gives spectral density with very narrow and deep minimum at frequency $2f_{\rm puls}$, which is evidently non-optimal for the NSNS and burst sources considered above. Moreover, technical noise makes it useless to have very deep minima in quantum noise spectral density, limiting $\rho_{\rm puls}$ by the value of
\begin{equation}
  \rho_{\rm puls}^{\rm max} \propto \frac{1}{\sqrt{S^h_{\rm tech}(4\pi f_{\rm puls})}} \,.
\end{equation}

Therefore, we optimize the following ``triple-purpose'' function:
\begin{equation}
  G_{\rm NS+burst+puls}(\lambda,\mu) = \left[
    \frac{1}{G_{\rm NS+burst}(\lambda)}
    + \mu\left(\frac{\rho_{\rm puls}^{\rm max}}{\rho_{\rm puls}}\right)^2
  \right]^{-1} \!\!\!,
\end{equation}
where $0\le\lambda\le 1$ and $\mu>0$ are Lagrange multipliers. We took pulsar J0034-0534 \cite{ATNF} as an example of millisecond pulsars, presumably emitting narrow-band high frequency GWs. Its barycentric rotational frequency is equal to $f_0\approx532.7\,{\rm Hz}$ and frequency of emitted GWs should be then $f_{\rm GW}=2f_0\approx1065.4\,{\rm Hz}$. Among the high-frequency pulsars, the distance to this one is significantly smaller compared to other ones ($0.98\,{\rm kpc}$) and therefore is one of the most probable candidates for GW detection. 

\begin{figure}[t]
  \begin{center} 
    \includegraphics[width=.49\textwidth]{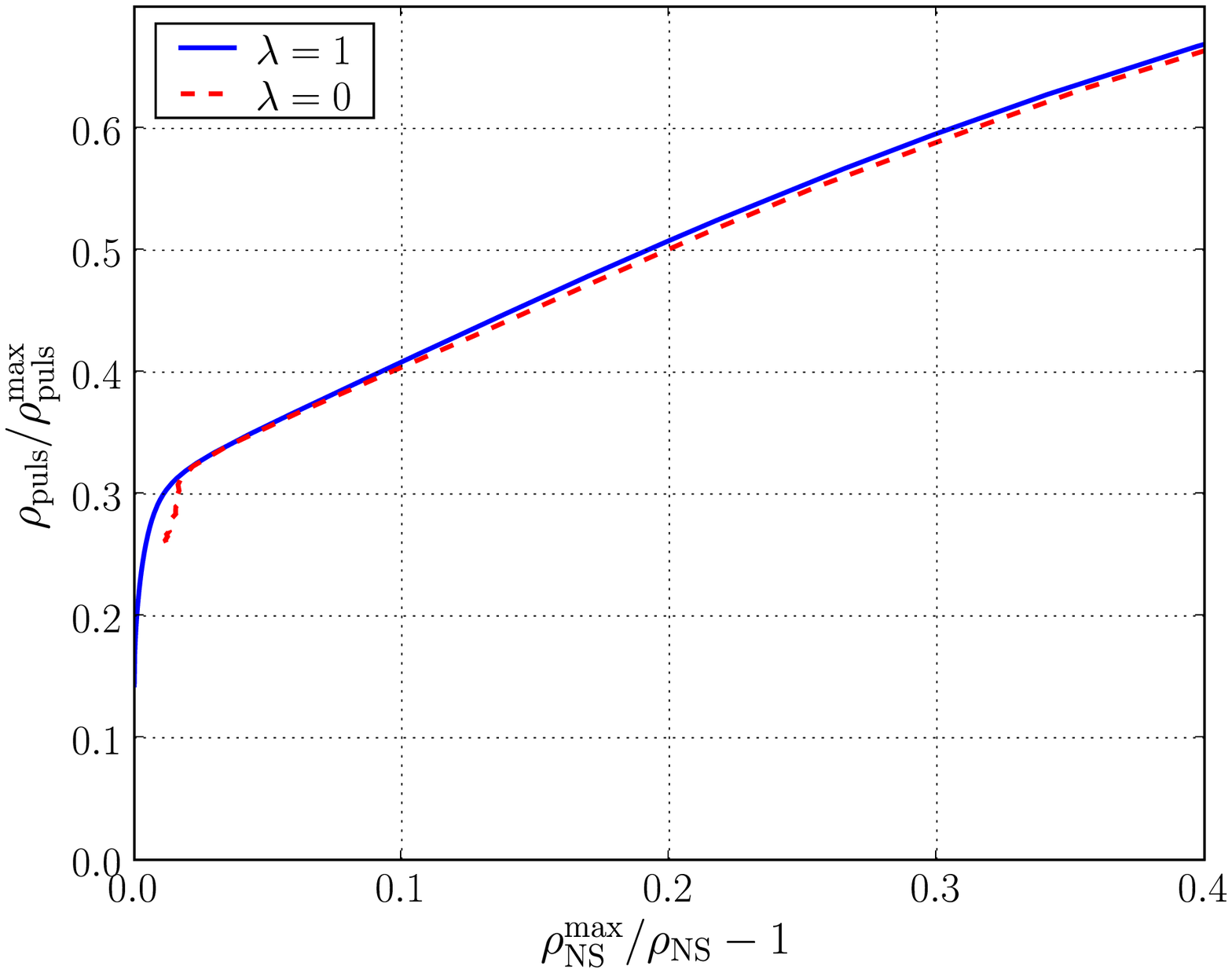}
    \includegraphics[width=.49\textwidth]{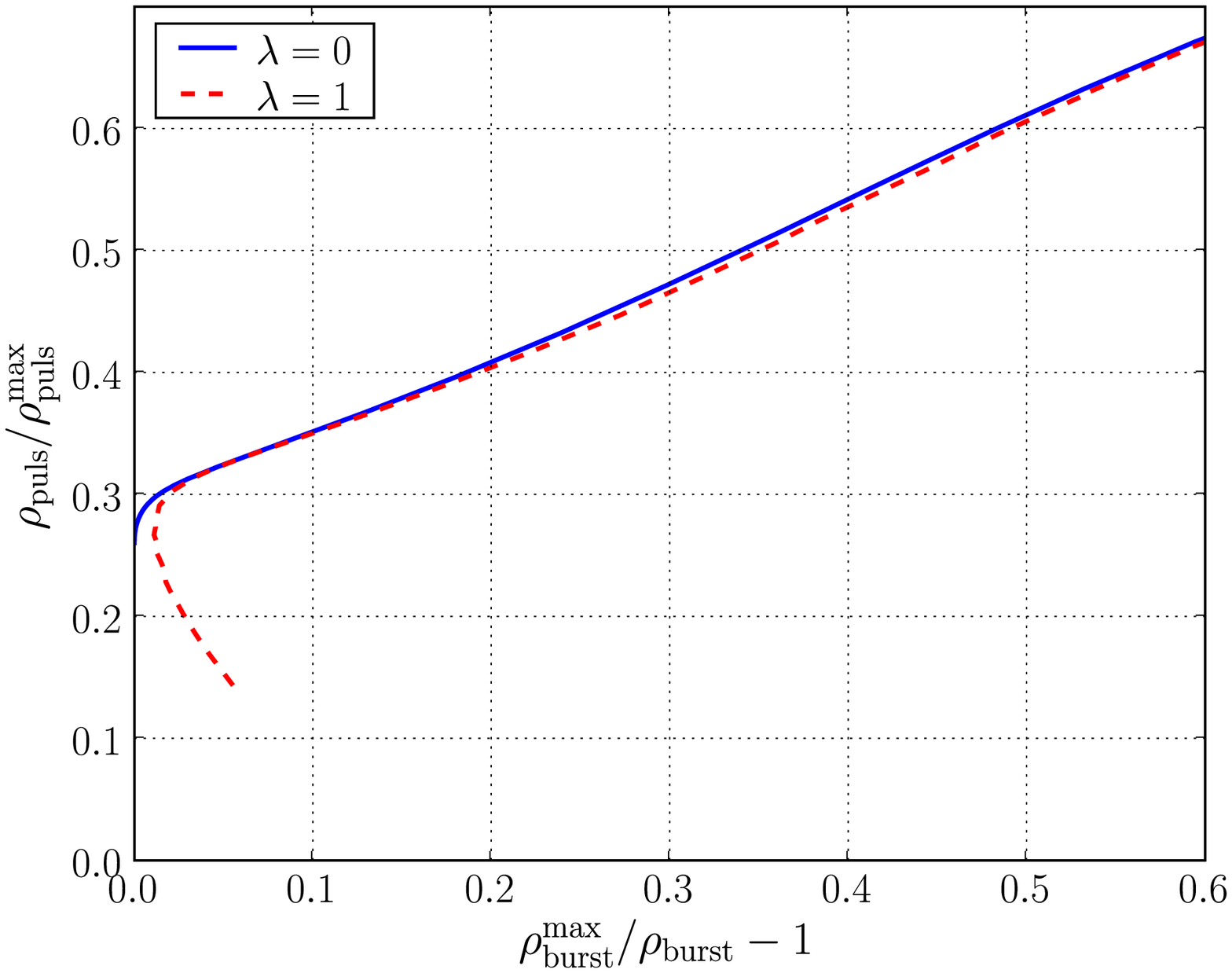}
  \end{center}
  \caption{Parametric plots of $\rho_{\rm NS}(\mu)/\rho_{\rm NS}^{\rm max}$ (top) and $\rho_{\rm burst}(\mu)/\rho_{\rm burst}^{\rm max}$ (bottom) against $\rho_{\rm puls}(\lambda)/\rho_{\rm puls}^{\rm max}$ for two optimizations regimes: with priority to $\rho_{\rm NS}$ ($\lambda=1$) and to $\rho_{\rm burst}$ ($\lambda=0$).} \label{fig:sco}
\end{figure}

The calculations results are presented in Fig.\,\ref{fig:nsns_burst} as parametric plots of $\rho_{\rm NS}(\mu)/\rho_{\rm NS}^{\rm max}$ and $\rho_{\rm burst}(\mu)/\rho_{\rm burst}^{\rm max}$ against $\rho_{\rm puls}(\mu)/\rho_{\rm puls}^{\rm max}$. 
It follows from these plots, that despite of two-dimensional character (two Lagrange parameters $\lambda$ and $\mu$) of the optimization procedure, the results are virtually one-dimensional, because only very small trade-off between values of $\rho_{\rm NS}$ and $\rho_{\rm burst}$ is possible (lines in both planes of Fig.\,\ref{fig:sco} almost coincide). 

\begin{figure}[b]
  \begin{center} 
    \includegraphics[width=.49\textwidth]{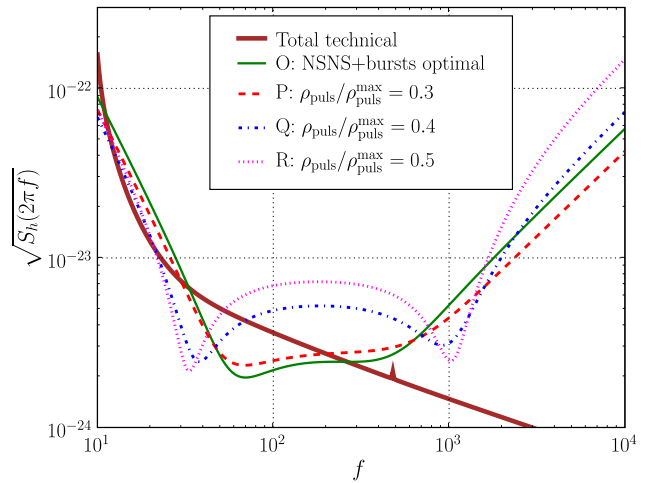}
  \end{center}
  \caption{Typical quantum noise spectral densities produced by the triple (NSNS+bursts+pulsars) optimization procedure}\label{fig:nsns_sco}
\end{figure}

In Fig.\,\ref{fig:nsns_sco} quantum noises for four characteristic regimes with optimal $G_{\rm NS+burst+puls}$ are plotted. These plots demonstate how the optimization algorithm increases the sensitivity at the given frequency $2f_{\rm puls}$. The starting point ``{\sf O}'' corresponds to obtained in Sec.\,\ref{Sec4} parameters set, optimized for NSNS and bursts signals, see Figs.\,\ref{fig4}, \ref{fig5}. First, the optimization algorithm tries to suppress the quantum noise in all high-frequency range ($f\gtrsim10^3\,{\rm Hz}$) by increasing $\varGamma$ and decreasing $\beta$, see point ``{\sf P}'' in these plots. At this stage, noticeable gain in pulsars sensitivity ($\sim 1.5$) can be obtained with negligibly small ($\sim1\%$) sensitivity loss for NSNS and bursts sources (see Table\,\ref{tab4.5}). Then, the optimization algorithm starts to ``grow'' local minimum at frequency $2f_{\rm puls}$, by increasing back $\beta$ in such a way that $\delta=\varGamma\sin\beta\to4\pi f_{\rm puls}$, see points ``Q'' and ``R''. At this stage $\rho_{\rm NS}$ and $\rho_{\rm burst}$ start to decrease noticeably (by tens of percents). However, $\rho_{\rm puls}$ increases several times at this stage.

\section{Conclusion}\label{Conc}

The results of this paper rely heavily on the estimates of the technical noise predicted for Advanced LIGO. These estimates almost definitely will be subject to changes during the next few years, however, it is improbable that technical noise estimates will change {\it significantly}. 
Therefore, all specific values obtained here should not be considered as final ones. 

The main result of this paper is not these values, but the conclusion that regimes of the signal-recycled interferometer exist which can provide good sensitivity for both binary inspiraling and burst gravitational wave sources. Moreover, ``triple-purposes'' regimes are also possible, which provide significant sensitivity gain for high-frequency periodical sources (millisecond pulsar) with only minor sensitivity degradation for binary inspiraling and bursts GW sources. 

Calculations presented in this paper show that in order to obtain good sensitivity for binary, burst and, to some extent, high-frequency periodic sources, it is necessary to use large values of interferometer bandwidth $\gamma\sim(2\div4)\times10^3\,{\rm s}^{-1}\gg2\times\pi100\,{\rm s}^{-1}$ with significant positive detunings $\delta\sim2\times10^3\,{\rm s}^{-1}$, see Table\,\ref{tab4.5}. These tunings give smooth broadband shape of quantum noise curves, dictated by the technical noises, especially by the mirrors thermal noise which has very flat frequency dependence in GW signals spectral range.

\acknowledgments

The work of F.~K. was supported in part by NSF and Caltech grant PHY-0353775. Work of I.~K. and D.~S. was supported by Russian government grant NSh-5178.2006.2. Work of S.~D. was supported by Alexander von Humboldt Foundation Research Fellowship.

Authors are grateful to Yanbei Chen for extremely useful suggestions and invaluable counselling, to Linqing Wen, Haixing Miao, Helge M\"uller-Ebhardt, Henning Rehbein and Kentaro Somiya for fruitful discussions and friendly encouragement. Special thanks to Rana Adhikari and all AdvLIGO Lab members for sharing their invaluable knowledge of specific features of real interferometers and many useful comments and suggestions that allowed us to improve this paper dramatically.  Authors also would like to thank MPI f\"ur Gravitationsphysik (AEI) both in Golm and in Hannover represented by directors Prof.~Dr.~B.~Schutz and Prof.~Dr.~K.~Danzmann for outstanding hospitality and cordial reception.

The paper has been assigned LIGO document number P080007-00-Z.

\end{document}